\documentclass[aps,twocolumn,superscriptaddress,prmaterials,10pt]{revtex4-2}

\usepackage{graphicx}
\usepackage{amsmath,amssymb}
\usepackage{bm}

\usepackage[colorlinks=true,linkcolor=blue!55!black,citecolor=blue!55!black,urlcolor=blue!55!black]{hyperref}
\usepackage{orcidlink}

\begin{document}

\title{Exchange topology and criticality in ferrite and chromium spinels:
a unified Monte Carlo analysis}

\author{Keltoum KHALLOUQ \orcidlink{0000-0002-5715-6351}}
\email{k.khallouq@gmail.com}
\affiliation{Laboratory of Solid Physics, Faculty of Sciences Dhar El Mahraz,
Sidi Mohamed Ben Abdellah University, Fez, Morocco}

\author{Ayoub EL MAAZOUZI \orcidlink{0000-0002-6181-5575}}
\email{ayb.elmaazouzi@gmail.com}
\affiliation{Laboratory of Solid Physics, Faculty of Sciences Dhar El Mahraz,
Sidi Mohamed Ben Abdellah University, Fez, Morocco}

\author{Kamal BOUMHARA \orcidlink{0000-0002-2722-4376}}
\email{kamalboumhara@gmail.com}
\affiliation{Laboratory of Solid Physics, Faculty of Sciences Dhar El Mahraz,
Sidi Mohamed Ben Abdellah University, Fez, Morocco}

\author{Rachid MASROUR \orcidlink{0000-0002-3646-665X}}  
\email{rachidmasrour@hotmail.com}
\affiliation{Laboratory of Solid Physics, Faculty of Sciences Dhar El Mahraz,
Sidi Mohamed Ben Abdellah University, Fez, Morocco}

\date{February 21, 2026}

\begin{abstract}
We report a unified analysis of Metropolis Monte Carlo results for two families
of magnetic spinels: inverse ferrites Fe$^{3+}_{A}$[M$^{2+}$Fe$^{3+}$]$_{B}$O$_4$
(M = Co, Cu, Fe, Ni), where superexchange couples two chemically distinct
sublattices, and chromium spinels $A$Cr$_2X_4$ ($A$ = Zn, Cd, Hg, $X$ = S, Se)
together with the breathing-lattice chromates Li$M$Cr$_4$O$_8$ ($M$ = Ga, In),
where a single Cr$^{3+}$ species occupies a corner-sharing tetrahedral network.
Placing the exchange constants, transition temperatures, critical exponents,
hysteresis, and magnetocaloric responses of these systems on a common footing,
we introduce two reduced quantities not previously reported: the ratio
$\theta_{\mathrm{CW}}/T_C$ for the ferrites and the normalized ordering scale
$t^{*}=k_BT_C/[J_1S(S+1)]$ for the chromium compounds. The ferrites cluster in
the range $\theta_{\mathrm{CW}}/T_C = 0.94$-$1.19$, close to the mean-field
expectation of unity and the signature of dominant, unfrustrated A-B
superexchange, and their exponents
($\beta = 0.20$-$0.26$, $\gamma = 1.23$-$1.27$, $\delta = 4.76$-$4.78$)
follow the three-dimensional Ising class. The chromium systems split into three
regimes: $t^{*} \approx 1.4$-$1.9$ for Ising-treated sulfides,
$t^{*} \approx 0.99$ for Heisenberg-treated selenides, and
$t^{*} \approx 0.24$-$0.25$ for the antiferromagnetic breathing chromates,
quantifying the combined suppression of $T_C$ by continuous spin symmetry and
by geometric frustration. Finite-thickness simulations of Fe$_3$O$_4$ resolve a
dimensionality crossover between two and four unit cells. We identify the
Ising-versus-Heisenberg dependence of the predicted universality class in
frustrated chromites as the principal open problem.
\end{abstract}

\maketitle

\section{Introduction}
\label{sec:intro}

Spinel compounds AB$_2X_4$ ($X$ = O, S, Se) crystallize in the cubic
$Fd\bar{3}m$ structure, with A cations on one-eighth of the tetrahedral
interstices and B cations on half of the octahedral interstices of a
close-packed anion lattice \cite{goodenough1958,kanamori1959}. Two chemically
distinct strategies place magnetic ions on this lattice, and they lead to
qualitatively different magnetism. In inverse ferrites
Fe$^{3+}_{A}$[M$^{2+}$Fe$^{3+}$]$_{B}$O$_4$, trivalent iron occupies the entire
tetrahedral sublattice while the octahedral sublattice hosts a mixture of
Fe$^{3+}$ and a divalent partner, so that superexchange proceeds along A-B,
A-A, and B-B pathways of markedly different strength, the near-180$^\circ$
A-O-B channel dominating \cite{kodama1997}. In chromium spinels
$A$Cr$_2X_4$ with nonmagnetic $A$ = Zn, Cd, Hg, the divalent cation merely
dilutes the lattice, the magnetic sublattice is carried entirely by
Cr$^{3+}$ ($3d^3$, $S = 3/2$) ions on the octahedral sites, which form a
network of corner-sharing tetrahedra intrinsically prone to geometric
frustration \cite{baltzer1966,ramirez1994}. The macroscopic consequence is a
gap of two orders of magnitude in ordering temperature: magnetite orders near
850-860 K \cite{kittel1996,levy2012}, whereas ZnCr$_2$S$_4$ orders below 16 K
and even the ferromagnetic selenide chromites remain confined to cryogenic
temperatures \cite{rudolf2007}.

Metropolis Monte Carlo simulation of Ising and Heisenberg Hamiltonians
parametrized by first-principles or experimentally fitted exchange constants
has become the standard route from this microscopic exchange topology to
critical temperatures, exponents, hysteresis, and magnetocaloric response
\cite{metropolis1953}. Over the past several years, our group has applied this approach
separately to the inverse ferrite series \cite{maazouzi2019a}, to bulk and
thin-film magnetite \cite{maazouzi2020a,maazouzi2020b}, to the sulfide
\cite{maazouzi2018} and selenide \cite{khallouq2023} chromites, to the
$A$-site-ordered chromates Li$M$Cr$_4$O$_8$ \cite{maazouzi2022a}, to the
spin-lattice-coupled ferrimagnet MnCr$_2$S$_4$ \cite{maazouzi2022b}, to the
cathode spinel LiMn$_{1.5}$Ni$_{0.5}$O$_4$ \cite{maazouzi2022c}, and to the
structurally related anisotropic pyrochlores $R_2M_2$O$_7$
\cite{khallouq2024}. Each of those studies addressed a single compound family
in isolation. What has been missing, in our work and in the wider simulation
literature on spinels, is a quantitative cross-comparison: the exchange
constants and ordering temperatures of ferrites and chromites have never been
reduced to a common dimensionless scale, so the respective contributions of
exchange strength, spin-model symmetry, and geometric frustration to the
ferrite-chromite temperature gap have remained entangled.

The present article closes that gap. We collect the exchange parameter sets,
Monte Carlo protocols, and simulated observables of the nine systems above,
recast them in a uniform notation, and extract two derived quantities that do
not appear in any of the source studies: the ratio of Curie-Weiss to critical
temperature, $\theta_{\mathrm{CW}}/T_C$, which measures the departure from
mean-field, unfrustrated behavior in the ferrites, and the normalized ordering
scale $t^{*} = k_BT_C/[J_1S(S+1)]$, which separates the roles of spin-model
symmetry and frustration in the chromium compounds. Section~\ref{sec:models}
defines the models and unifies the notation, Sec.~\ref{sec:method} summarizes
the simulation protocols and their differences, Sec.~\ref{sec:results}
presents the comparative results, including the new reduced-temperature
analysis, Sec.~\ref{sec:discussion} interprets the trends in terms of
superexchange pathways, frustration, dimensionality, and model choice, and
delimits the open questions.

\begin{figure}
\includegraphics[width=\columnwidth]{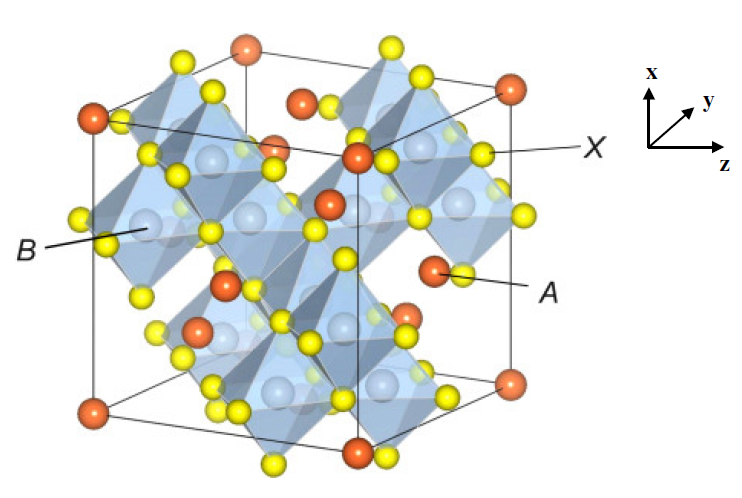}
\caption{Normal spinel structure AB2X4.}
\label{fig:structures}
\end{figure}

\section{Models and exchange parameter sets}
\label{sec:models}

Throughout this article we use $J^{(n)}$ ($n = 1$-$4$) for the $n$th-neighbor
exchange constant on a single magnetic sublattice, and $J_{\alpha\beta}$
($\alpha,\beta \in \{A, B_1, B_2\}$) for intersublattice couplings in
two-sublattice systems. All exchange constants are quoted in kelvin, where the
original parametrization is in meV we convert with
$1~\mathrm{meV} = 11.605$ K. Positive $J$ denotes ferromagnetic coupling in
the sign convention
$H = -\sum J_{ij}\,\bm{S}_i\!\cdot\!\bm{S}_j$ adopted here, the
antiferromagnetic convention of Ref.~\cite{maazouzi2022a} has been mapped onto
this one.

\subsection{Inverse ferrites}

For Fe$^{3+}_{A}$[M$^{2+}$Fe$^{3+}$]$_{B}$O$_4$ (M = Co, Cu, Fe, Ni), the
tetrahedral sublattice carries $S_A = 5/2$ throughout, while the octahedral
sublattice mixes $S_{B_1} = 5/2$ (Fe$^{3+}$) with an M-dependent
$S_{B_2}$ = 3/2, 1/2, 2, 1 for Co, Cu, Fe, Ni. The Ising Hamiltonian retains
six channels ($J_{AA}$, $J_{AB_1}$, $J_{AB_2}$, $J_{B_1B_1}$, $J_{B_1B_2}$,
$J_{B_2B_2}$), with values taken from the superexchange analysis of
Srivastava \emph{et al.} \cite{srivastava1979} and listed in
Table~\ref{tab:exchange}. Magnetite is the M = Fe member, for the dedicated
bulk- and thin-film-size studies a reduced three-channel description
($J_{AA} = -1.275$ K, $J_{BB} = +7.31$ K, $J_{AB} = -33.90$ K) from the
first-principles work of Uhl and Siberchicot \cite{uhl1995} was used.

\begin{table*}
\caption{Exchange parameter sets in unified notation (kelvin, positive =
ferromagnetic). Pyrochlore couplings, originally in meV, are converted with
1 meV = 11.605 K and carry an experimental uncertainty of $\approx$0.2 K
\cite{ross2011}. For Li$M$Cr$_4$O$_8$ the antiferromagnetic sign convention
of the source has been mapped onto the present one.}
\label{tab:exchange}
\begin{ruledtabular}
\footnotesize 9
\begin{tabular}{llll}
Compound & Channels & Values (K) & Parameter source \\
\hline
Fe$^{3+}$[Fe$^{3+}$Co$^{2+}$]O$_4$ & $J_{AA},J_{AB_1},J_{AB_2},J_{B_1B_1},J_{B_1B_2},J_{B_2B_2}$ & $-14,\,-28,\,-22.7,\,-9,\,-10.0,\,+46.0$ & \cite{srivastava1979} \\
Fe$^{3+}$[Fe$^{3+}$Cu$^{2+}$]O$_4$ & idem & $-14,\,-28,\,-27.0,\,-9,\,-10.0,\,+29.0$ & \cite{srivastava1979} \\
Fe$^{3+}$[Fe$^{3+}$Fe$^{2+}$]O$_4$ & idem & $-14,\,-28,\,-20.2,\,-9,\,-11.0,\,+44.0$ & \cite{srivastava1979} \\
Fe$^{3+}$[Fe$^{3+}$Ni$^{2+}$]O$_4$ & idem & $-14,\,-28,\,-27.4,\,-9,\,-10.0,\,+44.0$ & \cite{srivastava1979} \\
Fe$_3$O$_4$ (bulk/film studies) & $J_{AA},J_{BB},J_{AB}$ & $-1.275,\,+7.31,\,-33.90$ & \cite{uhl1995} \\
ZnCr$_2$S$_4$ & $J^{(1)},J^{(2)},J^{(3)}$ & $+2.20,\,-1.10,\,+0.55$ & \cite{hamedoun1986,hamedoun1995} \\
CdCr$_2$S$_4$ & idem & $+11.8,\,-0.44,\,-0.44$ & \cite{baltzer1965} \\
HgCr$_2$S$_4$ & idem & $+13.0,\,-0.70,\,0.00$ & \cite{baltzer1966} \\
CdCr$_2$Se$_4$ & $J^{(1)}$-$J^{(4)}$ & $+36.00,\,+0.25,\,-3.55,\,+0.45$ & \cite{yaresko2008} \\
HgCr$_2$Se$_4$ & idem & $+34.50,\,+0.05,\,-3.75,\,+0.45$ & \cite{yaresko2008} \\
LiGaCr$_4$O$_8$ & $J,\,J'$ & $-53.3,\,-31.98$ & \cite{okamoto2013} \\
LiInCr$_4$O$_8$ & $J,\,J'$ & $-56.8,\,-5.68$ & \cite{okamoto2013} \\
MnCr$_2$S$_4$ & $J_{\mathrm{Mn\text{-}Mn}},J_{\mathrm{Mn\text{-}Cr}},J_{\mathrm{Cr\text{-}Cr}}$ & $+3.4,\,+3.1,\,-9.1$ & \cite{miyata2020} \\
LiMn$_{1.5}$Ni$_{0.5}$O$_4$ ($U=0,3,5,7$ eV) & $J_{\mathrm{Mn\text{-}Mn}}$ & $-101,\,-142,\,-195,\,-310$ & \cite{maazouzi2022c} \\
 & $J_{\mathrm{Ni\text{-}Ni}}$ & $+764,\,+283,\,+115,\,+43.9$ & \cite{maazouzi2022c} \\
 & $J_{\mathrm{Mn\text{-}Ni}}$ & $+62.2,\,+50.6,\,+83.2,\,+81.6$ & \cite{maazouzi2022c} \\
Yb$_2$Ti$_2$O$_7$ & $J_1,J_2,J_3,J_4$ & $-1.04,\,-2.55,\,-3.37,\,+0.12$ & \cite{ross2011} \\
Er$_2$Ti$_2$O$_7$ & idem & $+1.28,\,-0.70,\,-1.16,\,-0.03$ & \cite{savary2012} \\
Er$_2$Sn$_2$O$_7$ & idem & $+0.81,\,+0.93,\,-1.28,\,+0.46$ & \cite{guitteny2013} \\
\end{tabular}
\end{ruledtabular}
\end{table*}

\subsection{Chromium spinels and chromates}

The sulfides ZnCr$_2$S$_4$, CdCr$_2$S$_4$, HgCr$_2$S$_4$ are described by an
Ising model with $J^{(1)}$-$J^{(3)}$ from Refs.~\cite{hamedoun1986,
hamedoun1995, baltzer1965}, the selenides CdCr$_2$Se$_4$ and HgCr$_2$Se$_4$
by a classical Heisenberg model with $J^{(1)}$-$J^{(4)}$ from the
band-structure analysis of Yaresko \cite{yaresko2008}. The chromates
Li$M$Cr$_4$O$_8$ ($M$ = Ga, In) realize a breathing pyrochlore lattice of
alternately expanded and contracted Cr$_4$ tetrahedra
($F\bar{4}3m$ symmetry), the two antiferromagnetic couplings $J$, $J'$ on the
two tetrahedron sizes are taken from the magnetization and NMR analysis of
Okamoto \emph{et al.} \cite{okamoto2013}, with breathing ratios
$J'/J = 0.60$ (Ga) and 0.10 (In). MnCr$_2$S$_4$ combines
$S_A = 5/2$ (Mn$^{2+}$) and $S_B = 3/2$ (Cr$^{3+}$) in a Heisenberg model
augmented by a spin-lattice displacement term with spring constant
$a' = 120$ \cite{miyata2020}. LiMn$_{1.5}$Ni$_{0.5}$O$_4$ places Mn ($S = 2$)
and Ni ($S = 3/2$) randomly on the octahedral sublattice, its three exchange
constants are obtained in this work's source study from DFT+$U$ total-energy
mapping over four collinear configurations, for $U = 0$-$7$ eV
\cite{maazouzi2022c}.

\subsection{Anisotropic pyrochlores}

The pyrochlore B sublattice of $R_2M_2$O$_7$ is topologically identical to the
Cr sublattice of the spinels, which motivates its inclusion here as a limiting
case of maximal exchange anisotropy. The bond-dependent Heisenberg model of
Ross \emph{et al.} \cite{ross2011} retains four couplings per bond
($J_1$: XY, $J_2$: Ising, $J_3$: symmetric off-diagonal, $J_4$:
Dzyaloshinskii-Moriya), with parameter sets fitted to inelastic neutron
scattering on Yb$_2$Ti$_2$O$_7$ \cite{ross2011}, Er$_2$Ti$_2$O$_7$
\cite{savary2012}, and Er$_2$Sn$_2$O$_7$ \cite{guitteny2013}, the quoted
experimental uncertainty on these couplings is 0.02 meV
($\approx 0.2$ K) \cite{ross2011}.

\section{Monte Carlo methodology}
\label{sec:method}

All simulations use single-spin-flip Metropolis dynamics
\cite{metropolis1953} with acceptance probability
$P = \exp(-\Delta E/k_BT)$, under periodic boundary conditions except for the
Fe$_3$O$_4$ thin-film geometry, where free boundaries are imposed along the
growth direction to expose surface effects. System sizes range from
$N = 1024$ spins (sulfide chromites) through $N = 3456$ ($16L^3$, $L = 6$,
selenide chromites and pyrochlores) to $N = 4096$ (ferrites) and
$N = 24L^3$ with $L$ up to 17 (magnetite size study). Sampling comprises
$5\times10^{5}$ Monte Carlo steps per spin with the first $5\times10^{4}$
discarded in the ferrite, chromite, chromate, and MnCr$_2$S$_4$ studies,
$10^{5}$ steps with $5\times10^{4}$ discarded for
LiMn$_{1.5}$Ni$_{0.5}$O$_4$, and a $10^{5}$-step equilibration followed by a
$10^{6}$-step measurement stage for the pyrochlores. Explicit equilibration
tests on magnetite show relaxation of the magnetization within
1400-2530 steps for $L = 11$-$17$ at $T = 100$ K, more than an order of
magnitude below the discard threshold \cite{maazouzi2020a}.

Magnetization, internal energy, susceptibility, and specific heat are
accumulated as thermal and fluctuation averages in the standard way, critical
temperatures are located from the susceptibility maximum, and critical
exponents from log-log fits to
$M \propto (T_C-T)^{\beta}$, $\chi \propto |T-T_C|^{-\gamma}$,
$M \propto H^{1/\delta}$, cross-checked where possible against the Rushbrooke
and Griffiths equalities \cite{rushbrook1963,griffiths1965}. The
Curie-Weiss temperature $\theta_{\mathrm{CW}}$ used in
Sec.~\ref{sec:reduced} is obtained from a linear fit of the simulated inverse
susceptibility $1/\chi(T)$ in the paramagnetic regime well above $T_C$, as
reported in Ref.~\cite{maazouzi2019a}, the fit range is not specified in that
source, a limitation noted in Sec.~\ref{sec:discussion}. Magnetocaloric
quantities follow from the Maxwell relation
$\Delta S_M(T,h) = \int_0^{h}(\partial M/\partial T)_{h'}\,dh'$ and the
relative cooling power
$\mathrm{RCP} = \int_{T_c}^{T_h}\Delta S_M\,dT$ over the full width at half
maximum. The source studies do not report statistical error bars on the
fitted exponents, nor the reduced-temperature window $|T-T_C|/T_C$ used for
the power-law fits, we return to the consequences of both omissions in
Sec.~\ref{sec:discussion}, and in what follows we treat exponent differences
smaller than $\pm 0.03$ as not resolved.

\section{Results}
\label{sec:results}

\subsection{Hierarchy of exchange constants}
\label{sec:hierarchy}

Table~\ref{tab:exchange} collects the exchange parameter sets in unified
notation. Two structural regularities emerge. First, wherever two chemically
distinct magnetic species coexist, the intersublattice channel dominates: in
the ferrites $|J_{AB_1}| = 28$ K exceeds $|J_{AA}| = 14$ K for every M, and in
magnetite $|J_{AB}| = 33.9$ K is five times $|J_{BB}|$ and more than twenty
times $|J_{AA}|$. The only ferromagnetic-sign coupling in the ferrite
Hamiltonians, $J_{B_2B_2}$, varies from $+29$ K (Cu) to $+46$ K (Co) but acts
between the minority spins and perturbs rather than competes with the A-B
channel. Second, in the single-sublattice chromium compounds the entire
exchange budget is confined to Cr-Cr pathways whose leading term is an order
of magnitude smaller: $J^{(1)} = +2.2$ to $+13.0$ K in the sulfides, rising to
$+34.5$-$+36.0$ K in the selenides, with subleading terms $J^{(2)}$,
$J^{(3)}$ that are frequently antiferromagnetic and, in the sulfides,
comparable in relative weight to $J^{(1)}$. The breathing chromates invert the
sign structure entirely: both $J$ and $J'$ are antiferromagnetic, at
53-57 K and 6-32 K respectively, so that every bond of every tetrahedron is
frustrated.

The DFT+$U$ mapping for LiMn$_{1.5}$Ni$_{0.5}$O$_4$ adds a caution relevant to
all first-principles-parametrized simulations in this field:
$J_{\mathrm{Ni\text{-}Ni}}$ collapses from 764 K at $U = 0$ to 43.9 K at
$U = 7$ eV while $|J_{\mathrm{Mn\text{-}Mn}}|$ triples over the same range
(Table~\ref{tab:exchange}), so the correlation strength assumed in the
electronic-structure step propagates directly into the classical spin model
and, through it, into every simulated observable.

\subsection{Transition temperatures on a common scale}
\label{sec:reduced}

Table~\ref{tab:tc} lists the simulated critical temperatures next to the
experimental and mean-field values quoted in the source studies. The raw
numbers reproduce the familiar picture: ferrites at 743-834 K within
2-5\% of experiment, sulfide chromites at 15.3-86 K, selenide chromites at
128.9-133.1 K, breathing chromates near 50 K, MnCr$_2$S$_4$ at 60 K, and the
anisotropic pyrochlores at 0.4 and 0.8 K for the two illustrative parameter
sets.

\begin{table}
\caption{Simulated critical temperatures alongside the experimental and
mean-field values quoted in the source studies. Experimental values are, in
several cases, secondary (handbook) values as cited by the source studies
rather than primary measurements, see Refs. in the last column for
provenance.}
\label{tab:tc}
\begin{ruledtabular}
\begin{tabular}{lccc}
Compound & $T_C^{\mathrm{MC}}$ (K) & $T_C^{\mathrm{exp}}$ (K) & Ref. \\
\hline
Fe$^{3+}$[Fe$^{3+}$Co$^{2+}$]O$_4$ & 830 & 793 & \cite{maazouzi2019a,smit1959} \\
Fe$^{3+}$[Fe$^{3+}$Cu$^{2+}$]O$_4$ & 743 & 728 & \cite{maazouzi2019a,smit1959} \\
Fe$^{3+}$[Fe$^{3+}$Fe$^{2+}$]O$_4$ & 834 & 858 & \cite{maazouzi2019a,smit1959} \\
Fe$^{3+}$[Fe$^{3+}$Ni$^{2+}$]O$_4$ & 834 & 858 & \cite{maazouzi2019a,smit1959} \\
Fe$_3$O$_4$ bulk ($L = 17$) & 815.14 & 850 & \cite{maazouzi2020a,kittel1996} \\
Fe$_3$O$_4$ film ($L = 1$-$4$) & 604-784 & - & \cite{maazouzi2020b} \\
ZnCr$_2$S$_4$ & 15.34 & 15.5 & \cite{maazouzi2018,hamedoun1986} \\
CdCr$_2$S$_4$ & 86.0 & 84.5 & \cite{maazouzi2018,baltzer1965} \\
HgCr$_2$S$_4$ & 66.07 & 60 & \cite{maazouzi2018,baltzer1966} \\
HgCr$_2$Se$_4$ & 128.86 & 106 & \cite{khallouq2023,lehmann1969} \\
CdCr$_2$Se$_4$ & 133.06 & 106 & \cite{khallouq2023,lehmann1969} \\
LiGaCr$_4$O$_8$ & $50 \pm 5$ & - & \cite{maazouzi2022a} \\
LiInCr$_4$O$_8$ & $50 \pm 5$ & - & \cite{maazouzi2022a} \\
MnCr$_2$S$_4$ & 60 & 64.5 & \cite{maazouzi2022b} \\
LiMn$_{1.5}$Ni$_{0.5}$O$_4$ & increases with $U$ & 113 & \cite{maazouzi2022c,singh2019} \\
Pyrochlore, conf.~1 & $T_N = 0.4$ & - & \cite{khallouq2024} \\
Pyrochlore, conf.~2 & $T_N = 0.8$ & - & \cite{khallouq2024} \\
\end{tabular}
\end{ruledtabular}
\end{table}

The comparative content of this article lies in Table~\ref{tab:reduced},
which reduces these temperatures to dimensionless form. For the ferrites we
form $\theta_{\mathrm{CW}}/T_C$ from the simulated Curie-Weiss and critical
temperatures of Ref.~\cite{maazouzi2019a}, the four compounds fall in the
narrow band 0.94-1.19. A ratio near unity is the mean-field signature of a
magnet whose dominant interactions are satisfied at the ordering transition,
values far above unity would indicate frustration-suppressed ordering
\cite{ramirez1994}. The ferrites therefore behave, on this measure, as
essentially unfrustrated ferrimagnets, notwithstanding the antiferromagnetic
sign of five of their six exchange channels: the bipartite A-B geometry
allows all dominant bonds to be satisfied simultaneously.

For the chromium compounds, where $\theta_{\mathrm{CW}}$ was not simulated
uniformly, we instead form the normalized ordering scale
\begin{equation}
t^{*} \,=\, \frac{k_B T_C}{J^{(1)}\,S(S+1)},
\qquad S = \tfrac{3}{2},
\label{eq:tstar}
\end{equation}
which measures the efficiency with which the leading exchange converts into
long-range order. Three regimes appear. The Ising-treated sulfides give
$t^{*} = 1.86$ (Zn), 1.94 (Cd), and 1.36 (Hg), the Heisenberg-treated
selenides give $t^{*} = 0.99$ (Cd) and 1.00 (Hg), and the breathing chromates,
evaluated with the mean coupling $\bar{J} = (J + J')/2$ to account for the two
tetrahedron sizes, give $t^{*} \approx 0.31$ (Ga) and 0.43 (In), or 0.25 and
0.23 when referred to the strong bond $J$ alone. The factor of roughly two
between the sulfide and selenide regimes is consistent with the well-known
reduction of $T_C$ by continuous spin symmetry: a Heisenberg magnet on the
same lattice with the same couplings orders at a substantially lower fraction
of $J S(S+1)$ than its Ising counterpart because transverse fluctuations
deplete the order parameter. The further factor of four separating the
chromates quantifies genuine geometric frustration: with all bonds
antiferromagnetic on corner-sharing tetrahedra, no configuration satisfies
the local constraints, and ordering is deferred to a small fraction of the
bare exchange scale. The anisotropic pyrochlores extend this sequence to its
extreme: with couplings of order 1-3 K (0.1-0.3 meV) the two studied
parameter sets order only at 0.4 and 0.8 K, and the factor of two between
them, produced by moving weight between the Ising ($J_2$) and off-diagonal
($J_3$) channels at fixed overall scale, demonstrates that in the strongly
frustrated limit the ordering temperature is controlled by the anisotropy
structure of the exchange matrix rather than by its magnitude.

Within the sulfide series itself, the ordering of $t^{*}$ values also carries
information. CdCr$_2$S$_4$, the only member whose second and third neighbor
couplings share the same (antiferromagnetic) sign, has the highest $t^{*}$,
HgCr$_2$S$_4$, with the largest $J^{(1)}$ but a vanishing $J^{(3)}$, has the
lowest. The subleading couplings therefore shift the ordering efficiency by
up to 40\% at fixed leading exchange, an effect invisible in the raw $T_C$
values, which are dominated by the fivefold spread of $J^{(1)}$.

\begin{figure*}
\includegraphics[width=0.8\textwidth]{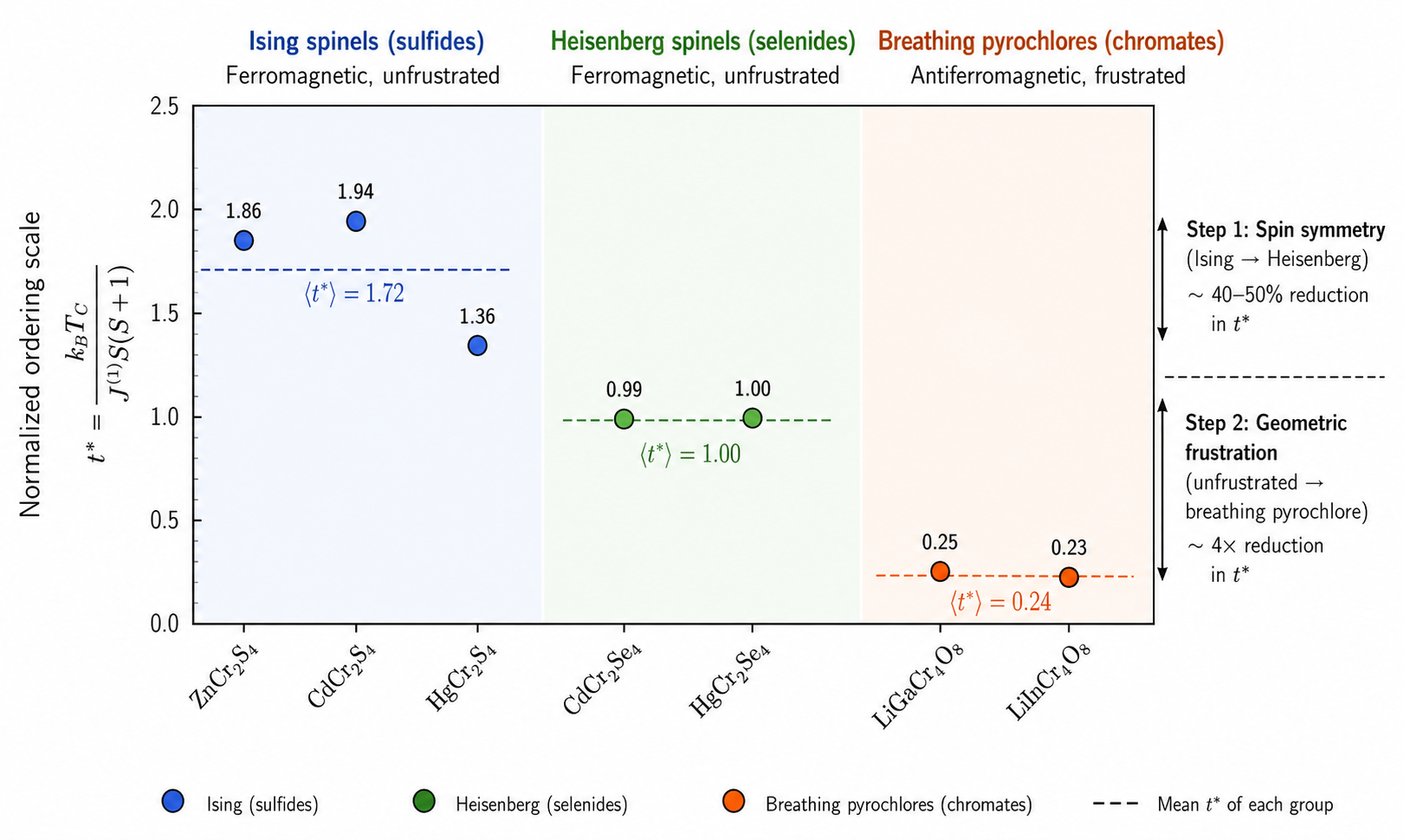}
\caption{Normalized ordering scale $t^{*} = k_BT_C/[J^{(1)}S(S+1)]$  for the chromium compounds listed in Table~\ref{tab:reduced}, grouped according to spin model and frustration regime. The decrease from the Ising sulfides to the Heisenberg selenides reflects the effect of spin symmetry, whereas the further reduction in the antiferromagnetic breathing chromates reflects the additional suppression of magnetic ordering by geometric frustration.}
\label{fig:tstar}
\end{figure*}

\subsection{Critical exponents and universality}
\label{sec:exponents}

\begin{table}
\caption{Fitted critical exponents (Ising-treated families) and selected
magnetocaloric and calculated transport observables (the MnCr$_2$S$_4$
thermoelectric entries combine FPLAPW electronic-structure calculations with
the Monte Carlo magnetic results and are not themselves Monte Carlo outputs).
Statistical uncertainties were not propagated in the source fits, differences
below $\pm 0.03$ are not regarded as significant here. Reference 3D Ising
values: $\beta = 0.22$, $\gamma = 1.24$, $\delta = 4.78$
\cite{leguillou1985,baillie1992,baker1978,george1984}.}
\label{tab:observables}
\begin{ruledtabular}
\begin{tabular}{lcccc}
Compound & $\beta$ & $\gamma$ & $\delta$ & Ref. \\
\hline
Fe$^{3+}$[Fe$^{3+}$Co$^{2+}$]O$_4$ & 0.202 & 1.234 & 4.765 & \cite{maazouzi2019a} \\
Fe$^{3+}$[Fe$^{3+}$Cu$^{2+}$]O$_4$ & 0.257 & 1.267 & 4.782 & \cite{maazouzi2019a} \\
Fe$^{3+}$[Fe$^{3+}$Fe$^{2+}$]O$_4$ & 0.212 & 1.233 & 4.764 & \cite{maazouzi2019a} \\
Fe$^{3+}$[Fe$^{3+}$Ni$^{2+}$]O$_4$ & 0.222 & 1.243 & 4.773 & \cite{maazouzi2019a} \\
ZnCr$_2$S$_4$ & 0.29 & 1.20 & 5.03 & \cite{maazouzi2018} \\
CdCr$_2$S$_4$ & 0.32 & 1.24 & 4.86 & \cite{maazouzi2018} \\
HgCr$_2$S$_4$ & 0.32 & 1.21 & 4.69 & \cite{maazouzi2018} \\
\hline
\multicolumn{5}{l}{Magnetocaloric / transport ($\mu_0h = 10$ T unless stated):} \\
\multicolumn{5}{l}{LiGaCr$_4$O$_8$: $-\Delta S_M^{\max} \approx 3.0$ J\,kg$^{-1}$K$^{-1}$, RCP $\approx 15$ J\,kg$^{-1}$ \cite{maazouzi2022a}} \\
\multicolumn{5}{l}{LiInCr$_4$O$_8$: $-\Delta S_M^{\max} \approx 1.6$ J\,kg$^{-1}$K$^{-1}$, RCP $\approx 26$ J\,kg$^{-1}$ \cite{maazouzi2022a}} \\
\multicolumn{5}{l}{MnCr$_2$S$_4$: $S_{\max} = 6.38\ \mu$V/K, $zT_{\max} \approx 1.85\times10^{-3}$ at 60 K \cite{maazouzi2022b}} \\
\end{tabular}
\end{ruledtabular}
\end{table}

\begin{table*}
\caption{Investigated compounds, structure types, magnetic sublattices, and
spin models. Original parametrizations and simulations are reported in the
references of the last column.}
\label{tab:classification}
\begin{ruledtabular}
\footnotesize 9
\begin{tabular}{llllll}
Compound & Structure & Space group & Magnetic sublattice(s) & Spin model & Ref. \\
\hline
Fe$^{3+}$[Fe$^{3+}$M$^{2+}$]O$_4$, M = Co/Cu/Fe/Ni & Inverse spinel & $Fd\bar{3}m$ & A: Fe$^{3+}$, B: Fe$^{3+}$/M$^{2+}$ & Ising & \cite{maazouzi2019a} \\
Fe$_3$O$_4$ (bulk, size study) & Inverse spinel & $Fd\bar{3}m$ & A: Fe$^{3+}$, B: Fe$^{3+}$/Fe$^{2+}$ & Ising & \cite{maazouzi2020a} \\
Fe$_3$O$_4$ (thin film) & Inverse spinel & $Fd\bar{3}m$ & idem & Ising & \cite{maazouzi2020b} \\
$A$Cr$_2$S$_4$, $A$ = Zn, Cd, Hg & Normal spinel & $Fd\bar{3}m$ & B: Cr$^{3+}$ & Ising & \cite{maazouzi2018} \\
$A$Cr$_2$Se$_4$, $A$ = Cd, Hg & Normal spinel & $Fd\bar{3}m$ & B: Cr$^{3+}$ & Heisenberg & \cite{khallouq2023} \\
Li$M$Cr$_4$O$_8$, $M$ = Ga, In & Breathing~(A-ordered) & $F\bar{4}3m$ & B: Cr$^{3+}$ & Ising ($S^z$) & \cite{maazouzi2022a} \\
MnCr$_2$S$_4$ & Normal spinel & $Fd\bar{3}m$ & A: Mn$^{2+}$, B: Cr$^{3+}$ & Heisenberg + lattice & \cite{maazouzi2022b} \\
LiMn$_{1.5}$Ni$_{0.5}$O$_4$ & Disordered spinel & $Fd\bar{3}m$ & B: Mn/Ni (random) & Ising & \cite{maazouzi2022c} \\
$R_2M_2$O$_7$, $R$ = Yb, Er, $M$ = Ti, Sn & Pyrochlore & $Fd\bar{3}m$ & Corner-sharing tetrahedra & Anisotropic Heisenberg & \cite{khallouq2024} \\
\end{tabular}
\end{ruledtabular}
\end{table*}

The fitted exponents (Table~\ref{tab:observables}) place both Ising-treated
families in the three-dimensional Ising class: the ferrites give
$\beta = 0.202$-$0.257$, $\gamma = 1.233$-$1.267$,
$\delta = 4.764$-$4.782$, and the sulfide chromites
$\beta = 0.29$-$0.32$, $\gamma = 1.20$-$1.24$, $\delta = 4.69$-$5.03$,
brackets that contain the accepted 3D Ising values
\cite{leguillou1985,baillie1992,baker1978,george1984}. For the sulfides the
specific-heat exponent obtained directly from scaling fits agrees with the
value reconstructed from the Rushbrooke equality
$\alpha = 2 - \gamma - 2\beta$ to within 0.02-0.03 (e.g.,
ZnCr$_2$S$_4$: 0.18 direct versus 0.20 reconstructed), an internal consistency
check that the ferrite study did not perform. Given the absence of quoted
statistical uncertainties and the modest lattice sizes
(Sec.~\ref{sec:method}), we do not regard the residual spread within either
family as physically significant, what the data establish is compatibility
with, not high-precision confirmation of, 3D Ising universality. No exponent
extraction exists for the Heisenberg-treated selenides, MnCr$_2$S$_4$, or the
pyrochlores, a gap whose implications are taken up in
Sec.~\ref{sec:discussion}.

\subsection{Finite-size and finite-thickness behavior of magnetite}
\label{sec:size}

The dedicated size study of bulk magnetite \cite{maazouzi2020a} yields
$T_C(L) = 813.40$, 814.31, 814.51, 814.93, 815.00, 815.14 K for
$L = 7$-$17$, a monotonic rise of 0.2\% consistent in sign and form with the
standard finite-size correction $T_C(\infty) - T_C(L) \propto L^{-1/\nu}$,
converging toward 815-817 K, in agreement with the earlier simulational
estimate of 817 K \cite{mazozuluaga2004} and below the experimental 850 K
\cite{kittel1996}. The thin-film geometry behaves very differently:
$T_C = 604$, 717, 750, 784 K for thicknesses of one to four unit cells
\cite{maazouzi2020b}, a 26\% suppression at monolayer thickness that decays
rapidly and crosses over, above roughly six unit cells, to a slow asymptotic
approach to the bulk value near 816 K. The contrast between the two data
sets, plotted together against inverse size in Fig.~\ref{fig:fss}, cleanly
separates the ordinary finite-size correction of a three-dimensional system
from a genuine dimensionality crossover, and localizes the latter between two
and four unit cells, consistent with the reported surface transition
temperature of 723 K for the magnetite (100) surface \cite{bartelt2013}.
Coercivity and remanence follow the same logic: in the film they grow with
thickness and decay with temperature, and in the bulk they decay smoothly
from $H_C = 0.13$ T, $M_r = 86.4$ emu/g at 100 K to
$H_C = 0.001$ T, $M_r = 38.9$ emu/g at 800 K \cite{maazouzi2020a}.

\subsection{Hysteresis systematics and magnetocaloric response}
\label{sec:hysteresis}

All ferrite and chromite compounds show saturated hysteresis loops below
$T_C$ whose coercive field, remanence, and saturation magnetization decrease
monotonically with temperature, collapsing into reversible
superparamagnetic-like curves above the transition. Two chemical trends cut
across the families. In the ferrites, saturation magnetization and coercivity
decrease with increasing $3d$ electron count of the divalent octahedral
cation \cite{maazouzi2019a}, in the sulfide chromites, the low-temperature
coercive field orders as
$h_C(\mathrm{Hg}) > h_C(\mathrm{Cd}) > h_C(\mathrm{Zn})$, tracking the atomic
number of the nonmagnetic A cation \cite{maazouzi2018}. Both trends point to
the same conclusion: the hysteresis scale in these simulations is set by the
effective exchange field of the reversing sublattice rather than by any
explicit anisotropy term, since none is present in the Hamiltonians.

The frustrated compounds add qualitatively new field-driven physics. The
breathing chromates and the selenide chromites both display multiple
hysteresis branches at low temperature, merging into a single loop as
$T \to T_C$ \cite{maazouzi2022a,khallouq2023}, and MnCr$_2$S$_4$ exhibits a
robust half-magnetization plateau between $h \approx 25$-30 and 80-90 T in
which the Cr moments align while the antiparallel Mn pair contributes zero
net moment, the simulated phase-boundary fields (21 and 89 T at 4 K, drifting
to 25 and 81 T at 17 K) reproduce the anomalies observed experimentally up to
60 T \cite{miyata2020,maazouzi2022b}. The spin-lattice term is essential to
this plateau: the simulations show the Cr sublattice relaxing to zero
displacement within the plateau even when all species are allowed to move,
stabilizing the collinear phase against the noncollinear Yafet-Kittel
alternative \cite{yafet1952}.

Magnetocaloric characterization exists for the chromates and selenides. In
both, $-\Delta S_M(T)$ peaks at $T_C$ and grows with field, peak values read
from the simulated entropy curves at $\mu_0h = 10$ T are of order
1.6 J\,kg$^{-1}$K$^{-1}$ for LiInCr$_4$O$_8$ and
3.0 J\,kg$^{-1}$K$^{-1}$ for LiGaCr$_4$O$_8$, with corresponding relative
cooling powers of approximately 26 and 15 J\,kg$^{-1}$
\cite{maazouzi2022a}. In the selenides, both $-\Delta S_M$ and the adiabatic
temperature change oscillate below 100 K, a direct thermodynamic echo of the
frustration-induced magnetization plateaus, and the RCP of HgCr$_2$Se$_4$
peaks near 140 K, close to its transition \cite{khallouq2023}. The larger
entropy change of the Ga chromate correlates with its larger breathing ratio
$J'/J = 0.60$: the more nearly uniform pyrochlore limit carries more
field-releasable entropy near the transition than the strongly dimerized In
compound ($J'/J = 0.10$), whose small tetrahedra are already strongly
correlated well above $T_C$.

\begin{table}
\caption{Reduced ordering scales derived in this work from the data of
Tables~\ref{tab:exchange} and \ref{tab:tc}. For the ferrites,
$\theta_{\mathrm{CW}}$ is the simulated Curie-Weiss temperature of
Ref.~\cite{maazouzi2019a}. For the chromium compounds,
$t^{*} = k_BT_C/[J^{(1)}S(S+1)]$ with $S = 3/2$, for the chromates $t^{*}$ is
given relative to the strong bond $|J|$, with the value relative to
$|\bar{J}| = (|J|+|J'|)/2$ in parentheses. Uncertainties on $t^{*}$ are
dominated by the $\pm 5$ K uncertainty on $T_C$ quoted for the chromates and
are of order $\pm 10\%$, no statistical errors were propagated for $T_C$ in
the other source studies, so no uncertainty is quoted for those entries.}
\label{tab:reduced}
\begin{ruledtabular}
\begin{tabular}{lcc}
Compound & $\theta_{\mathrm{CW}}/T_C$ & $t^{*}$ \\
\hline
Fe$^{3+}$[Fe$^{3+}$Co$^{2+}$]O$_4$ & 1.07 & - \\
Fe$^{3+}$[Fe$^{3+}$Cu$^{2+}$]O$_4$ & 1.19 & - \\
Fe$^{3+}$[Fe$^{3+}$Fe$^{2+}$]O$_4$ & 0.94 & - \\
Fe$^{3+}$[Fe$^{3+}$Ni$^{2+}$]O$_4$ & 1.15 & - \\
ZnCr$_2$S$_4$ (Ising) & - & 1.86 \\
CdCr$_2$S$_4$ (Ising) & - & 1.94 \\
HgCr$_2$S$_4$ (Ising) & - & 1.36 \\
CdCr$_2$Se$_4$ (Heisenberg) & - & 0.99 \\
HgCr$_2$Se$_4$ (Heisenberg) & - & 1.00 \\
LiGaCr$_4$O$_8$ & - & 0.25 (0.31) \\
LiInCr$_4$O$_8$ & - & 0.23 (0.43) \\
\end{tabular}
\end{ruledtabular}
\end{table}

\section{Discussion}
\label{sec:discussion}

\subsection{Why the ferrite-chromite gap is topological, not energetic}

A naive reading of Table~\ref{tab:tc} would attribute the two-order-of-magnitude
gap between ferrite and chromite ordering temperatures to weaker exchange in
the chromites. Table~\ref{tab:reduced} shows that this is at most half the
story. The selenide chromites carry a leading exchange
($J^{(1)} \approx 35$ K) comparable to the dominant ferrite channel
($|J_{AB_1}| = 28$ K), yet order at 130 K rather than 800 K. The decisive
difference is the spin-length weighting and the connectivity: in the ferrites
the dominant bond couples two $S = 5/2$ ions (or $S = 5/2$ to
$S \geq 1$) across a bipartite A-B geometry in which every strong bond can
be satisfied, so the effective ordering energy
$\sim |J_{AB}|\,S_A S_B$ reaches several hundred kelvin per bond, in the
chromites the same bare $J$ couples two $S = 3/2$ ions on a lattice where the
antiferromagnetic subleading terms cannot all be satisfied. The
$\theta_{\mathrm{CW}}/T_C$ ratios of 0.94-1.19 for the ferrites, a spread of
at most 19\% about the mean-field value of unity, and the progression
$t^{*} \approx 1.9 \to 1.0 \to 0.25$ from Ising sulfides through
Heisenberg selenides to fully frustrated chromates, decompose the gap
quantitatively into an exchange-and-spin-length factor, a spin-symmetry
factor of roughly two, and a frustration factor of roughly four.

\subsection{Superexchange pathways and the Goodenough-Kanamori frame}

The sign pattern of Table~\ref{tab:exchange} is consistent throughout with
the Goodenough-Kanamori rules \cite{goodenough1958,kanamori1959}. The
near-180$^\circ$ Fe$_A$-O-Fe$_B$ pathway between half-filled $d^5$
configurations is strongly antiferromagnetic and anchors the ferrimagnetism
of every ferrite studied, its uniformity across M explains why the four
compounds share $J_{AA}$, $J_{AB_1}$, and $J_{B_1B_1}$ identically and differ
only through the M-dependent channels. In the chromites the near-90$^\circ$
Cr-$X$-Cr geometry produces a ferromagnetic $J^{(1)}$ whose strength grows
with anion polarizability (S $\to$ Se) and with the lattice expansion imposed
by the A cation, while the longer superexchange paths through two anions
generate the antiferromagnetic $J^{(2)}$, $J^{(3)}$ responsible for
frustration. The chromates push this to the limit: contraction of half the
tetrahedra strengthens direct antiferromagnetic Cr-Cr overlap, flipping even
the nearest-neighbor coupling to antiferromagnetic and converting the lattice
into a breathing analog of the pyrochlore antiferromagnet.

\subsection{Site ordering, correlation strength, and control of the exchange landscape}

Two of the systems function as controlled perturbations of the chromite
template. In Li$M$Cr$_4$O$_8$, the $A$-site cation order tunes the breathing
ratio $J'/J$ by a factor of six between the Ga and In members while leaving
$T_C$ near 50 K in both, showing that in the fully frustrated regime the
ordering temperature is pinned by the average exchange while the breathing
anisotropy instead reshapes the field-induced plateau structure and the
magnetocaloric weight (Sec.~\ref{sec:hysteresis}). In
LiMn$_{1.5}$Ni$_{0.5}$O$_4$, the Hubbard $U$ plays the analogous role on the
electronic side: the simulated magnetization curves change qualitatively
between $U = 3$, 5, and 7 eV, and the transition temperature rises with $U$
toward the experimental 113 K
\cite{maazouzi2022c,singh2019,hanafusa2016}. Taken together, the two systems
show that cation order and correlation strength are largely independent
levers on the exchange landscape, one geometric and one electronic, and that
simulated observables inherit the full sensitivity of whichever lever is
uncertain.

\subsection{Model choice, universality, and the limits of the present data}

The recovery of 3D Ising exponents in both Ising-treated families
(Sec.~\ref{sec:exponents}) is expected on universality grounds: short-range
Ising models on three-dimensional lattices share a single critical class
irrespective of lattice decoration \cite{leguillou1985}. The substantive
question is therefore not whether the fits recover Ising values but whether
the Ising description is physically justified for each material. For the
ferrites, whose ground states are collinear and whose dominant couplings are
strong, the approximation is defensible. For the chromites the situation is
less comfortable: the same family has been treated with Ising spins (sulfides)
and Heisenberg spins (selenides), and Eq.~(\ref{eq:tstar}) shows that this
choice alone moves the normalized ordering temperature by a factor of two.
A magnet correctly described by weakly anisotropic Heisenberg spins would
exhibit 3D Heisenberg exponents over most of the critical region, crossing
over to Ising values only asymptotically close to $T_C$ if a small easy-axis
anisotropy is present, the exponent sets O($n$) field theory assigns to the
two classes differ well outside the spread of the present fits
\cite{leguillou1985, Pelissetto2002}. Because no exponent extraction was performed for the
Heisenberg-treated systems, the simulation record assembled here cannot yet
discriminate between these scenarios, and we regard a uniform
Ising-versus-Heisenberg comparison at fixed exchange topology, with quoted
statistical errors and a Binder-cumulant \cite{Binder1981} or finite-size-scaling determination
of $\nu$, as the single most valuable next computation in this program.

Three further limitations bound the conclusions. First, statistical
uncertainties on the fitted exponents were not propagated in the source
studies, and the lattice sizes ($L \leq 17$, $N \leq 4096$-$3456$) are modest
by the standards of dedicated critical-phenomena work, the exponent
comparisons in Sec.~\ref{sec:exponents} should accordingly be read as
consistency checks, not precision tests. Second, dipolar interactions are
omitted throughout, although in the chromites and pyrochlores the exchange
scale is small enough that dipolar corrections to the low-field hysteresis
cannot be excluded a priori \cite{Fennell2009}. Third, the exchange inputs mix first-principles,
experimental-fit, and DFT+$U$ provenances, and the pyrochlore couplings carry
a quoted uncertainty of 0.2 K on couplings of order 1-3 K
\cite{ross2011,khallouq2024}, the factor-of-two spread in pyrochlore
$T_N$ between the two parameter sets studied is a fair measure of the
resulting model uncertainty in the strongly frustrated limit.

\section{Conclusion}
\label{sec:conclusion}

Reducing the Monte Carlo record for nine spinel and pyrochlore systems to a
common dimensionless scale separates, for the first time within a single
framework, the three mechanisms behind the ferrite-chromite ordering gap.
Ferrites order at $\theta_{\mathrm{CW}}/T_C \approx 1$ because their dominant
A-B superexchange is bipartite and unfrustrated, chromium spinels lose a
factor of about two in normalized ordering temperature to continuous spin
symmetry and a further factor of about four to geometric frustration when all
tetrahedral bonds turn antiferromagnetic, with the anisotropic pyrochlores
marking the limit in which the exchange-matrix structure, not its magnitude,
controls $T_N$. Cation ordering (breathing ratio) and electronic correlation
strength (Hubbard $U$) act as independent controls on this landscape,
reshaping plateaus and magnetocaloric weight at nearly fixed $T_C$ in the
first case and shifting $T_C$ itself in the second. Finite-thickness
simulations of magnetite localize its two- to three-dimensional crossover
between two and four unit cells. The clearest open problem exposed by the
comparison is methodological: the predicted universality class of the
frustrated chromites currently depends on an untested choice between Ising
and Heisenberg spin models, and settling it requires a uniform two-model
simulation with controlled statistics on the same exchange topology.

\bibliographystyle{apsrev4-2}   
\bibliography{bib}

\end{document}